\documentclass[12pt,a4paper]{panl}
\usepackage[english]{babel}
\usepackage{cite}
\usepackage{wrapfig}
\usepackage{graphicx}
\usepackage{amssymb}
\usepackage{amsfonts}
\usepackage{amsmath}
\usepackage{longtable}
\usepackage{rotating}
\usepackage{lscape}
\usepackage{epsfig}
\usepackage{multirow}

\usepackage{soul}

\originalTeX
\begin{document}

\title{DD-pair production in the parton Reggeization approach taking into account single and double parton scattering scenarios}
\maketitle
\authors{L.\,Alimov$^{a,}$\footnote{E-mail: alimov.le@yandex.ru},
V.\,Saleev$^{a,b,}$\footnote{E-mail: saleev.vladimir@gmail.com}}

\from{$^{a}$\,Samara National Research University}
\from{$^{b}$\,Joint Institute for Nuclear Research}

\begin{abstract}

The paper presents the results of calculations of the cross sections for the production of $DD$-pairs at centre-of-mass energies $\sqrt{s}=7$ and $13$ TeV in the parton Reggeization approach. The contribution of both single and double parton scattering is taken into account.
To describe the non-perturbative effects of $c$-quark hadronization into $D$-mesons, a fragmentation model with the Peterson fragmentation function is used.
The parameter $\sigma_{eff}=11.5^{+0.6}_{-0.5}$ mb, which determines the contribution of the double parton scattering mechanism, is fixed on the basis of a comparison with the available LHCb collaboration experimental data on the $D$-meson pairs production at $\sqrt{s}=7$ TeV.
\end{abstract}
\vspace*{6pt}

\noindent
PACS: 13.85.$-$t; 14.40.Be 

\label{sec:intro}
\section*{Introduction}

Theoretical and experimental studies of multiple particle production processes at high energies at the Large Hadron Collider (LHC) explicitly point to the necessity of taking into account the contribution of multiple parton scattering, primarily double parton scattering (DPS) \cite{Diehl:2017wew, Shao:2024ehw}.

One of the most effective methods for calculating cross sections of multiple particle production is the high-energy factorization approach, or $k_T$-factorization. In this approach, leading contributions from additional parton emissions are resummed and included into unintegrated parton distribution functions (uPDFs). One can restrict the calculation of parton-level cross sections to leading order of perturbation theory in the strong coupling constant.

At the same time, it is necessary to take into account the contribution of processes with several additional partons in the final state in calculations within a fixed order of perturbation theory in the collinear parton model, which is a non-trivial task even when studying inclusive particle production.

The contribution of single parton scattering (SPS) is insufficient to describe the existing experimental data on the production of $D$-meson pairs \cite{LHCb:2012aiv} as was shown in Ref.~\cite{vanHameren:2015wva}.
Taking into account the gluon fragmentation mechanism allows one to obtain a reasonably good description of the experimental data \cite{Maciula:2016wci}. However, this raises a number of issues, such as double counting of various contributions and the inability to apply the massless gluon fragmentation model to $D$-mesons in the small transverse momenta region $p_T \le m_D$ \cite{Karpishkov:2016gda}.

In this paper we discuss the results of calculations of various differential cross sections for the production of $D^0$-meson pairs in proton collisions within the Parton Reggeization Approach (PRA) \cite{Nefedov:2013ywa, Karpishkov:2017umm, Nefedov:2020ugj}, which is a gauge-invariant scheme of the $k_T$-factorization approach based on the effective QCD action at high energies by L.~N. Lipatov \cite{Lipatov:1995pn}.
The calculations were performed at LHC energies of $\sqrt{s}=7$ and $\sqrt{s}=13$ TeV, in two rapidity regions: $2<y<4$ and $|y|<2$. The contributions from SPS and DPS were taken into account, and the DPS model parameter $\sigma_{eff}$ was obtained by fitting the LHCb experimental data \cite{LHCb:2012aiv} at $\sqrt{s}=7$ TeV. A comparison of the calculation results with the predictions obtained in Ref.~\cite{vanHameren:2015wva} is also presented.
Previously, within the PRA, the processes of pair production of $\Upsilon D$ \cite{Karpishkov:2019vyt}, $J/\psi J/\psi$ \cite{Chernyshev:2022gek}, $J/\psi D$ \cite{Chernyshev:2023qea}, $J/\psi \gamma$ \cite{Alimov:2024pqt} were studied, also taking into account SPS and DPS.

\label{sec:PRA}
\section*{Parton Reggeization Approach}

At high energies, a gauge-invariant version of the $k_T$-factorization approach based on the factorization model of the cross section in the multi-Regge kinematic limit, known as the Parton Reggeization Approach (PRA), is used to describe the production cross sections of hard processes \cite{Nefedov:2013ywa, Karpishkov:2017umm, Nefedov:2020ugj}.

In general form, the production cross section of a $DD$-pair in the process $pp\to DDX$ can be expressed as a convolution:
\begin{align}
\label{eq:factorization}
    \sigma^{PRA}(pp\to DDX)=\sum\limits_{a,b}\int\frac{dx_1}{x_1}\frac{d^2q_{1T}}{\pi}\int\frac{dx_2}{x_2}\frac{d^2q_{2T}}{\pi}\times
\end{align}
\begin{align*}
    \times\Phi_a(x_1,q_{1T}^2,\mu^2)\Phi_b(x_2,q_{2T}^2,\mu^2)\hat{\sigma}^{PRA}(ab\to DD)
\end{align*}
where the summation is done over all types of Reggeized partons $a,b = R, Q, \bar{Q}$.

In the modified Kimber–Martin–Ryskin–Watt (KMRW) model, the uPDFs $\Phi_{a,b}(x,q^2_T,\mu^2)$ are normalized to the collinear PDFs \cite{Nefedov:2020ugj}:
\begin{align}
\label{eq:normalization}
\int\limits_0^{\mu^2}dq_T^2\Phi_a(x,q^2_T,\mu^2)=xf_a(x,\mu^2)
\end{align}
In contrast to the original KMRW model, the Sudakov form factor in the improved model \cite{Nefedov:2020ugj} depends on $x$:
\begin{align}
    T_a(t,\mu^2,x)=\exp\left[ -\int\limits_t^{\mu^2}\frac{dt'}{t'} \frac{\alpha_S(t')}{2\pi}\left( \tau_a(t',\mu^2)+\Delta\tau_a(t',\mu^2,x) \right) \right]
\end{align}
where $F_{a,b}(x,t)=x f_{a,b}(x,t)$ and the KMR cutoff functions $\Delta(t,\mu^2)=\mu/(\mu+\sqrt{t})$ \cite{Kimber:2001sc}:
\begin{align}
\tau_a(t,\mu^2)=\sum\limits_b\int\limits_{0}^{1}dz zP_{ba}(z)\theta(\Delta(t,\mu^2)-z)
\end{align}
\begin{align}
\Delta\tau_a(t,\mu^2,x)=\sum\limits_b\int\limits_0^1 dz\theta(z-\Delta(t,\mu^2))\left[zP_{ba}-\frac{F_b(\frac{x}{z},t)}{F_a(x,t)}P_{ab}(z)\theta(z-x) \right]
\end{align}

The improved KMRW uPDFs resum the contributions of Sudakov logarithms $\log(\mu/\sqrt{t})$ in quark and gluon uPDFs and can be used to describe processes with arbitrary $x$, not only in $x \ll 1$ region \cite{Nefedov:2020ugj}.

The cross sections of parton subprocesses are expressed in terms of the squared matrix elements $\overline{|M|^2}_{\text{PRA}}$, calculated according to the Feynman rules of Lipatov's effective theory \cite{Lipatov:1995pn} in the standard way:
\begin{align}
d\hat{\sigma}^{\text{PRA}}(ab\to c\bar{c})=(2\pi)^4\delta^{(4)}(q_1+q_2-p_1-p_2)\frac{\overline{|M|^2}_{\text{PRA}}}{2x_1x_2s} d\Phi
\end{align}
where $d\Phi$  is the phase space of the final-state particles.


\label{sec:fragmentation}
\section*{Fragmentation model}

Hadronization of $c$-quarks into $D$-mesons is described in the fragmentation model by means of nonperturbative fragmentation functions (FFs). The production cross section of $D$-mesons is expressed as a convolution of the FFs with the production cross section of $c$-quarks:
\begin{align}
\label{eq:fragmentation}
    d\sigma(pp\to DX)=P_{c\to D}\times \int\limits_{z^{min}}^{1} dz  D_{c\to D}(z)  d\sigma(pp\to c\bar{c} X)
\end{align}
where $z^{min}=m_{D}/(E_c+|\vec{p}_c|)$, $P_{c\to D}$ is the hadronization probability of $c$-quark into $D$-meson. For the $D^0$-meson, the probability obtained from experimental data is $P_{c\to D^0}=0.542$ \cite{Gladilin:1999nf}. In calculations taking into account the masses of $c$-quarks and $D$-mesons, there is an uncertainty in the choice of the hadronization parameter $z$. In this work, it was assumed that $m_D=1.87$ GeV/c${}^2$, $m_c=1.3$ GeV/c${}^2$ and
\begin{align}
    z=\frac{E_{D}+|\vec{p}|_{D}}{E_{c}+|\vec{p}|_{c}}.
\end{align}

We use the hard-scale-independent Peterson FF, with $\vec p_D/|\vec p_D|=\vec p_c/|\vec p_c|$:
\begin{align}
    D_{c\to D}(z)=\frac{N}{z(1-\frac{1}{z}-\frac{\epsilon}{1-z})^2}
\end{align}
where $\epsilon=0.06$, $N$ is the $D_{c\to D^0}(z)$ normalization factor.

\label{sec:d-production}
\section*{Pair $D$-meson production}

Within the SPS mechanism, taking into account (\ref{eq:factorization}) and (\ref{eq:fragmentation}), the total production cross section of a $DD$ pair is written as a convolution of two uPDFs, two fragmentation functions, and the parton-level cross section of the process $ab \to cc \bar c\bar c$ 
\begin{align}
\label{eq:sps-eq}
    \sigma^{\text{SPS}}(pp\to DDX)=P^2_{c\to D} \sum\limits_{a,b}   D_{c\to D}(z_1)\otimes D_{c\to D}(z_2) \otimes
\end{align}
\begin{align*}
    \otimes  \Phi_a \otimes \Phi_b \otimes \hat{\sigma}^{\text{PRA}}(ab\to c\bar{c}c\bar{c})
\end{align*}
We take into account gluon fusion $RR\to c\bar{c}c\bar{c}$ and quark--antiquark annihilation $Q\bar{Q}\to c\bar{c}c\bar{c}$ processes.

The total production cross section of a $DD$-pair in DPS is expressed through the product of the total production cross sections of single $D$-mesons \cite{Calucci:1999yz}:
\begin{align}
\label{eq:dps-eq}
    \sigma^{\text{DPS}}=\frac{\sigma^{\text{SPS}}(pp\to D X)\times \sigma^{\text{SPS}}(pp\to D X)}{n\cdot\sigma_{eff}}
\end{align}
where $n=2$ for identical particles and $n=1$ for different ones, $\sigma_{eff}$ is the phenomenological parameter that determines the contribution of DPS. In the case of DPS, the following subprocesses are taken into account: gluon fusion $RR\to c\bar{c}\;\otimes\;RR\to c\bar{c}$, quark--antiquark annihilation $Q\bar{Q}\to c\bar{c}\;\otimes\;Q\bar{Q}\to c\bar{c}$ and mixed subprocesses $RR\to c\bar{c}\;\otimes\;Q\bar{Q}\to c\bar{c}$.

\label{sec:kaite}
\section*{Calculation methods}

If the analytic form of the matrix elements $\overline{|M|^2}_{\text{PRA}}$ is known, then the calculation of the multidimensional integrals in (\ref{eq:sps-eq}) and (\ref{eq:dps-eq}) is carried out numerically using Monte Carlo methods.
However, for $2\to 4$ processes, obtaining an analytic formula for $\overline{|M|^2}_{\text{PRA}}$ is difficult due to the large number of diagrams, and the resulting expressions are too cumbersome for efficient multidimensional integration with high precision.
An alternative is to use parton-level generators based on the Monte Carlo method.
In this work, the KaTie Monte Carlo generator \cite{vanHameren:2016kkz} is used, which implements the spinor-helicity formalism and a recursive method for calculating tree amplitudes based on the BCFW relations \cite{Britto:2005fq}.
It was previously shown \cite{Nefedov:2013ywa, Kutak:2016mik} that the results of calculations of amplitudes for various processes performed in KaTie coincide with the results obtained using the Feynman rules of the effective theory of L.~N. Lipatov.
As the collinear PDFs required for constructing the improved KMRW uPDFs, the MSTW2008lo parametrization was used.

\label{sec:results}
\section*{Results}

Calculations of the differential cross sections for $D^0D^0$-pair production at $\sqrt{s}=7$ TeV in the large-rapidity region (Fig. 1) have been performed. Fitting the experimental data of the LHCb collaboration \cite{LHCb:2012aiv} gives $\sigma_{eff}=11.5^{+0.6}_{-0.5}$ mb, which is in good agreement with the value of $\sigma_{eff}$ obtained in the PRA for the processes of $J/\psi J/\psi$ \cite{Chernyshev:2022gek, Chernyshev:2024nei} and $J/\psi D$ \cite{ Chernyshev:2023qea} pair production. Hereafter, the differential cross sections are calculated at the hard scale $\mu=(m_{T1}^c+m_{T2}^c)/2$, where $m_T=\sqrt{m_c^2+p_{Tc}^2}$. The theoretical uncertainty of the calculations arising from varying the hard scale by a factor of 2 is shown for the total differential cross section in the PRA as a gray band.

The differences between the differential cross sections in the PRA and the results of \cite{vanHameren:2015wva} presented in Fig. 1 are due not only to the differences in the uPDFs used, but also to the choice of the $c$-quark mass. In this work, $m_c=1.3$ GeV/c${}^2$ is used, whereas in \cite{vanHameren:2015wva} $m_c=1.5$ GeV/c${}^2$ is used. The cross sections in the PRA turn out to be larger, which is consistent with the ratio of the $m_c$ values used and the choice of hard scales.

Using the obtained value of $\sigma_{eff}$, we predict the differential cross sections for $D^0D^0$ pair production as functions of the pair invariant mass $M$, the absolute rapidity difference $|\Delta y|$, the azimuthal angle difference $|\Delta\phi|/\pi$, and the transverse momentum of a single $D$-meson $p_T$ at the LHC energy $\sqrt{s}=13$ TeV in the rapidity regions $2<y<4$ and $|y|<2$, see Fig. 2.

At the considered energies, $7$ TeV and $13$ TeV, a dominance of the DPS contribution characteristic of multiparticle production processes is observed.


\section*{Acknowledgments}
The work is supported by the Foundation for the Advancement of Theoretical Physics and Mathematics BASIS, grant No. 24–1–1–16–5 and by the grant of the Ministry of Science and Higher Education of the Russian Federation, No. FSSS-2025–0003


\begin{figure}[th]
\begin{center}
    \begin{minipage}[b]{0.48\textwidth}
    \includegraphics[width=\textwidth]{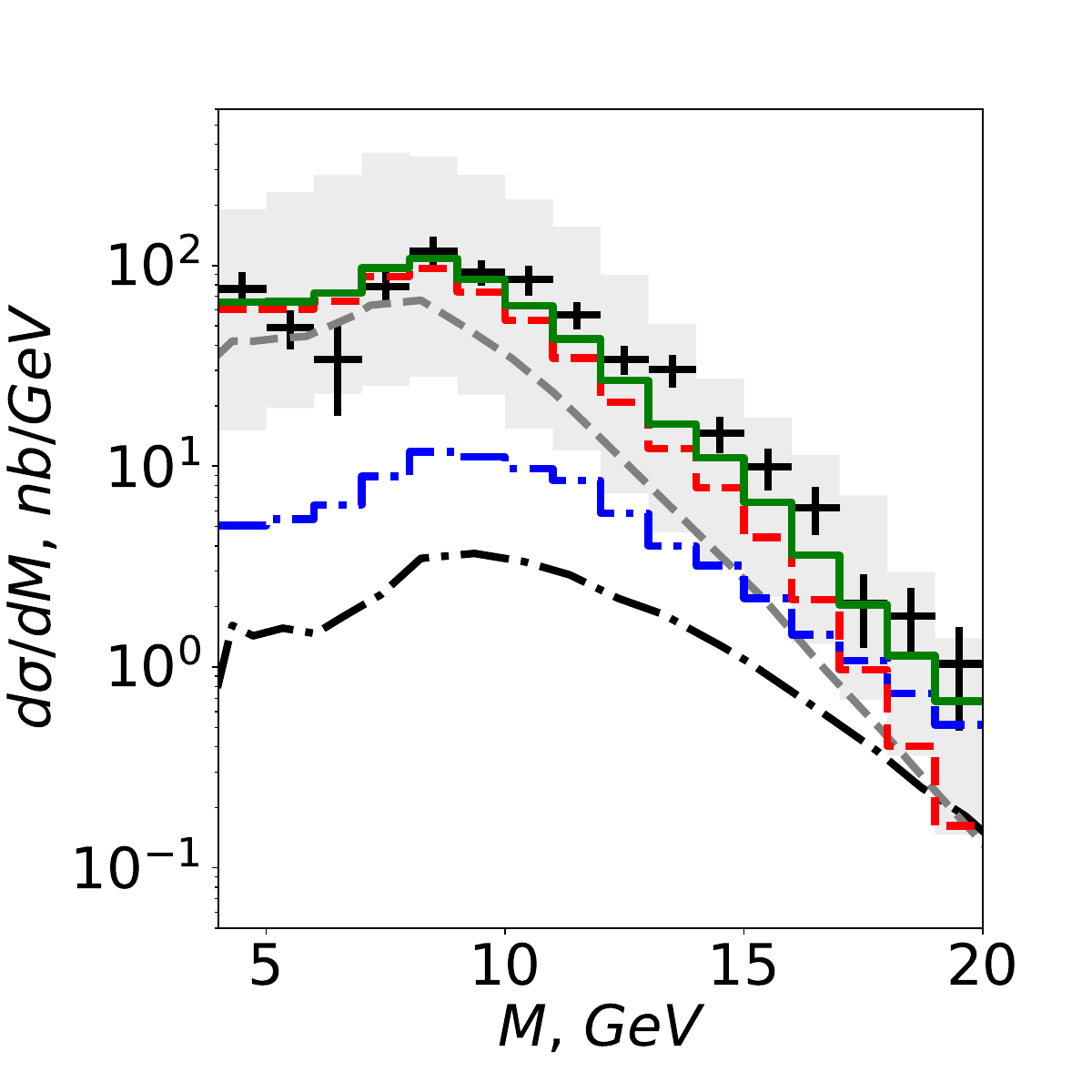}
    \end{minipage}
    \begin{minipage}[b]{0.48\textwidth}
    \includegraphics[width=\textwidth]{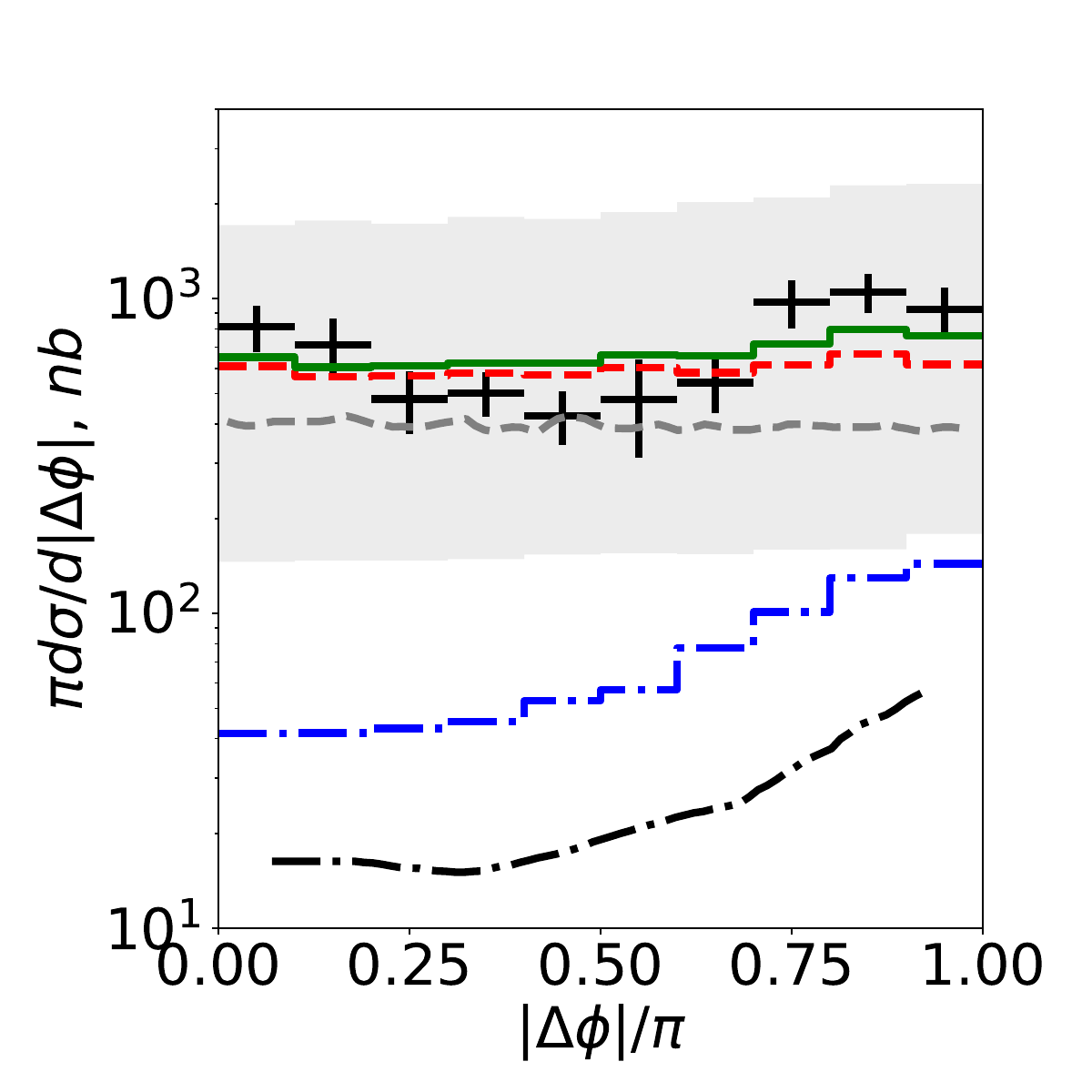}
    \end{minipage}
\vspace{-3mm}
\label{pic:test_kt}

\caption{Differential cross sections for $D^0D^0$-pair production as functions of the pair invariant mass $M$ (left) and the azimuthal angle difference $|\Delta\phi|/\pi$ (right) at $\sqrt{s}=7$ TeV, $2<y<4$ and $p_{T}^{D}>3$ GeV/c. The SPS contribution is shown by the blue dash-dotted curve, the DPS contribution by the red dashed curve, and the total production cross section in the PRA by the solid green curve. The SPS contribution from \cite{vanHameren:2015wva} is shown by the black dash-dotted curve, and the DPS contribution from \cite{vanHameren:2015wva} by the gray dashed curve. Experimental data from the LHCb collaboration \cite{LHCb:2012aiv}.}
\end{center}
\labelf{fig01}
\vspace{-5mm}
\end{figure}

\begin{figure}[t]
\begin{center}
    \begin{minipage}[b]{0.4\textwidth}
    \includegraphics[width=\textwidth]{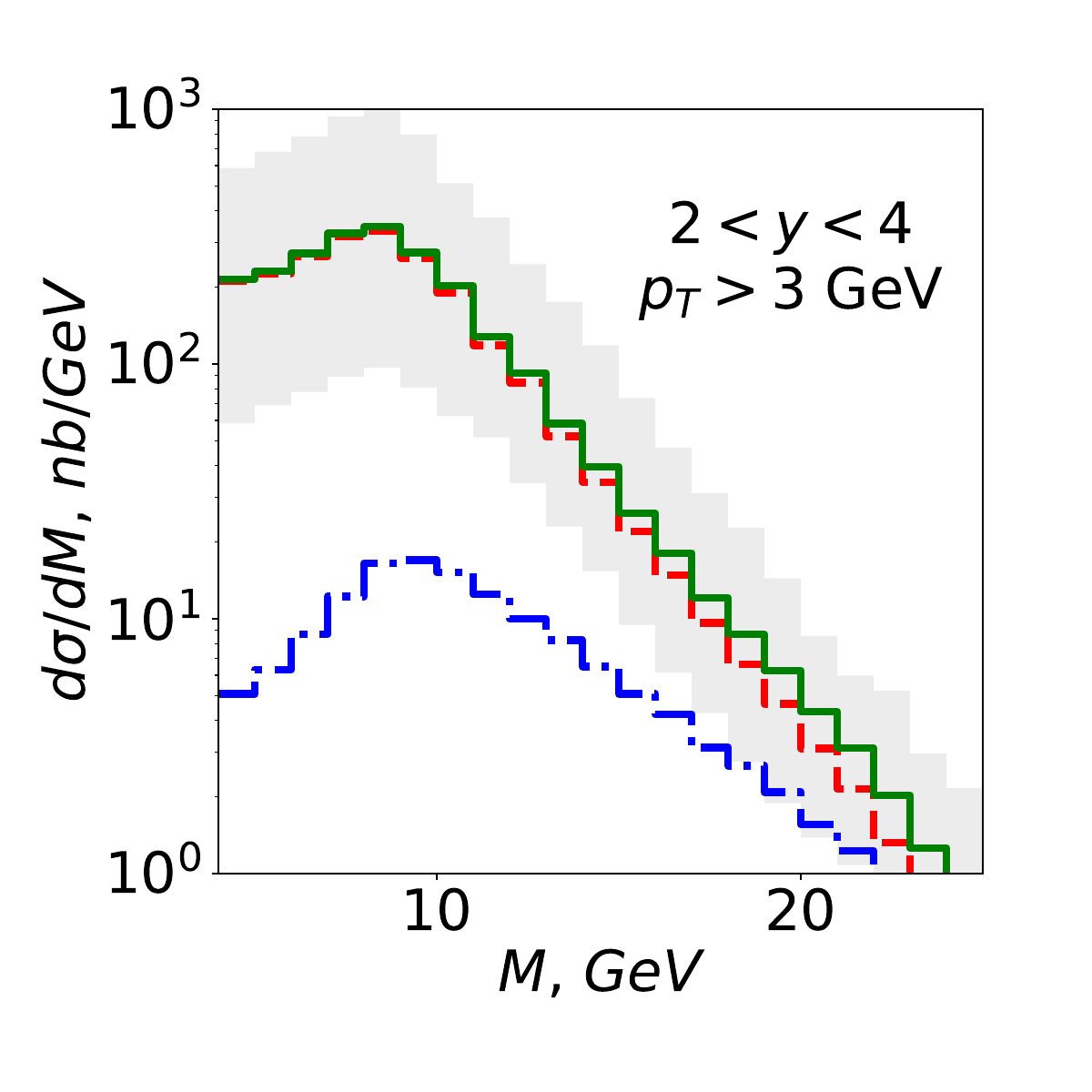}
    \vspace{-10mm}
    \end{minipage}
    \begin{minipage}[b]{0.4\textwidth}
    \includegraphics[width=\textwidth]{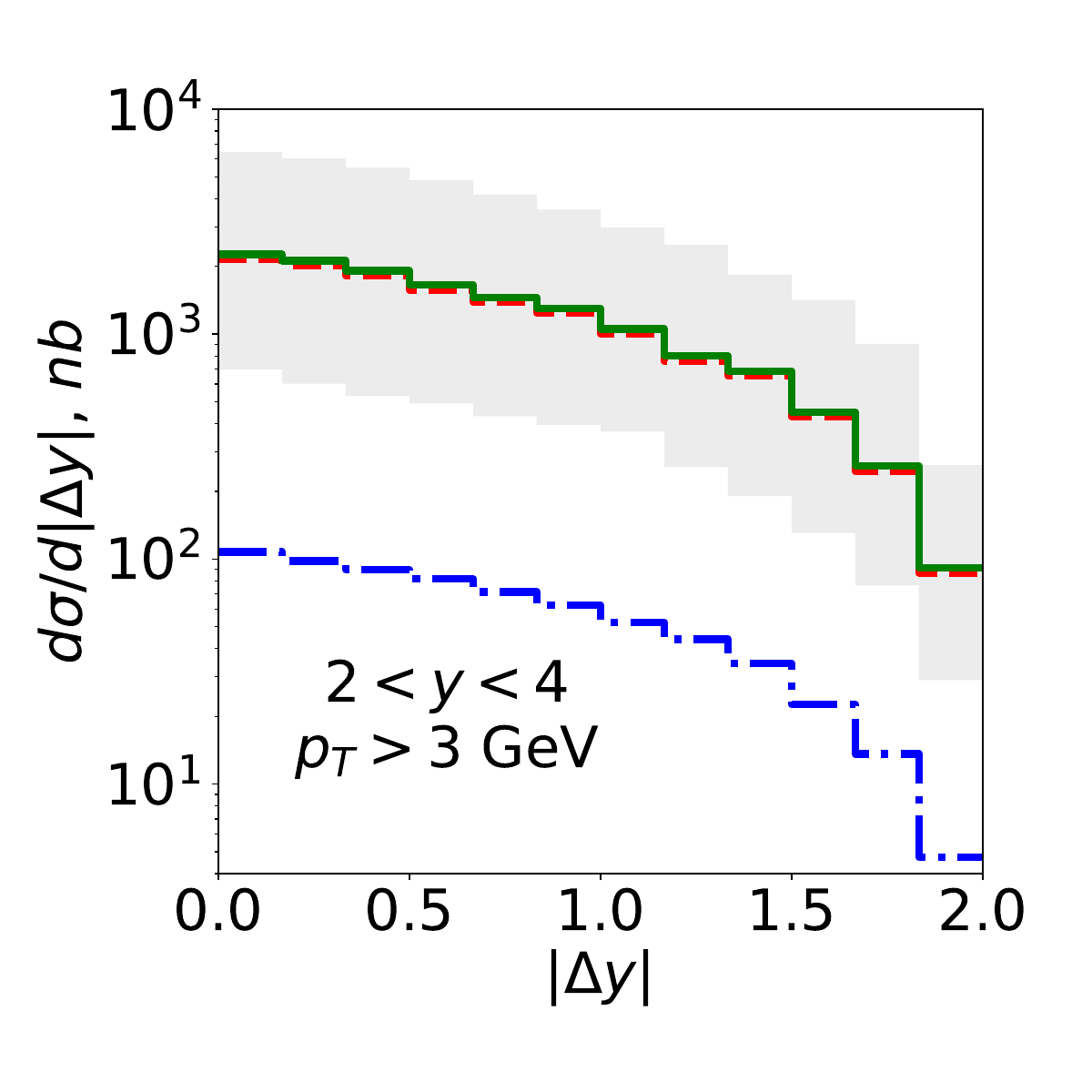}
    \vspace{-10mm}
    \end{minipage}
    \vspace{-5mm}
    \begin{minipage}[b]{0.4\textwidth}
    \includegraphics[width=\textwidth]{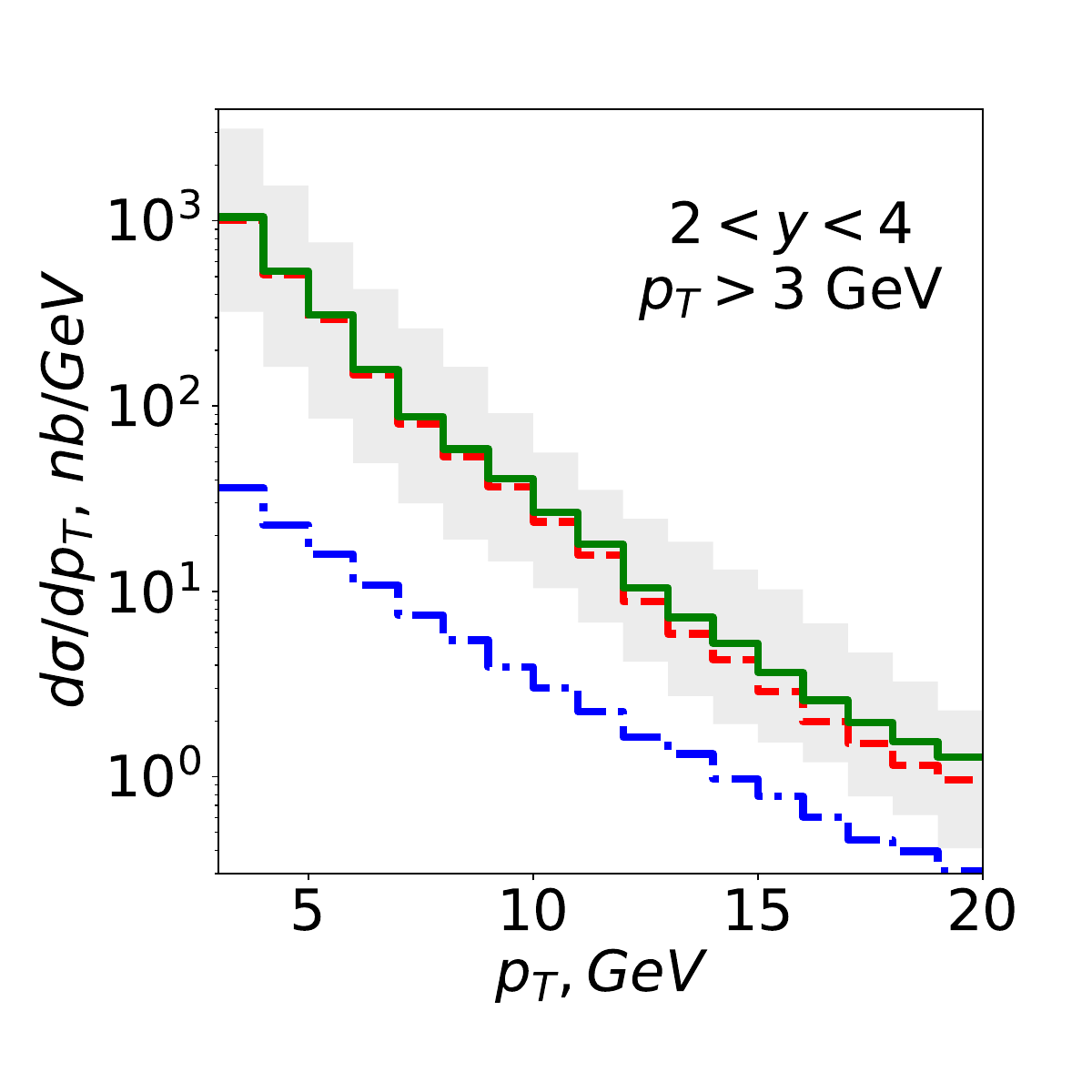}
    \end{minipage}
    \begin{minipage}[b]{0.4\textwidth}
    \includegraphics[width=\textwidth]{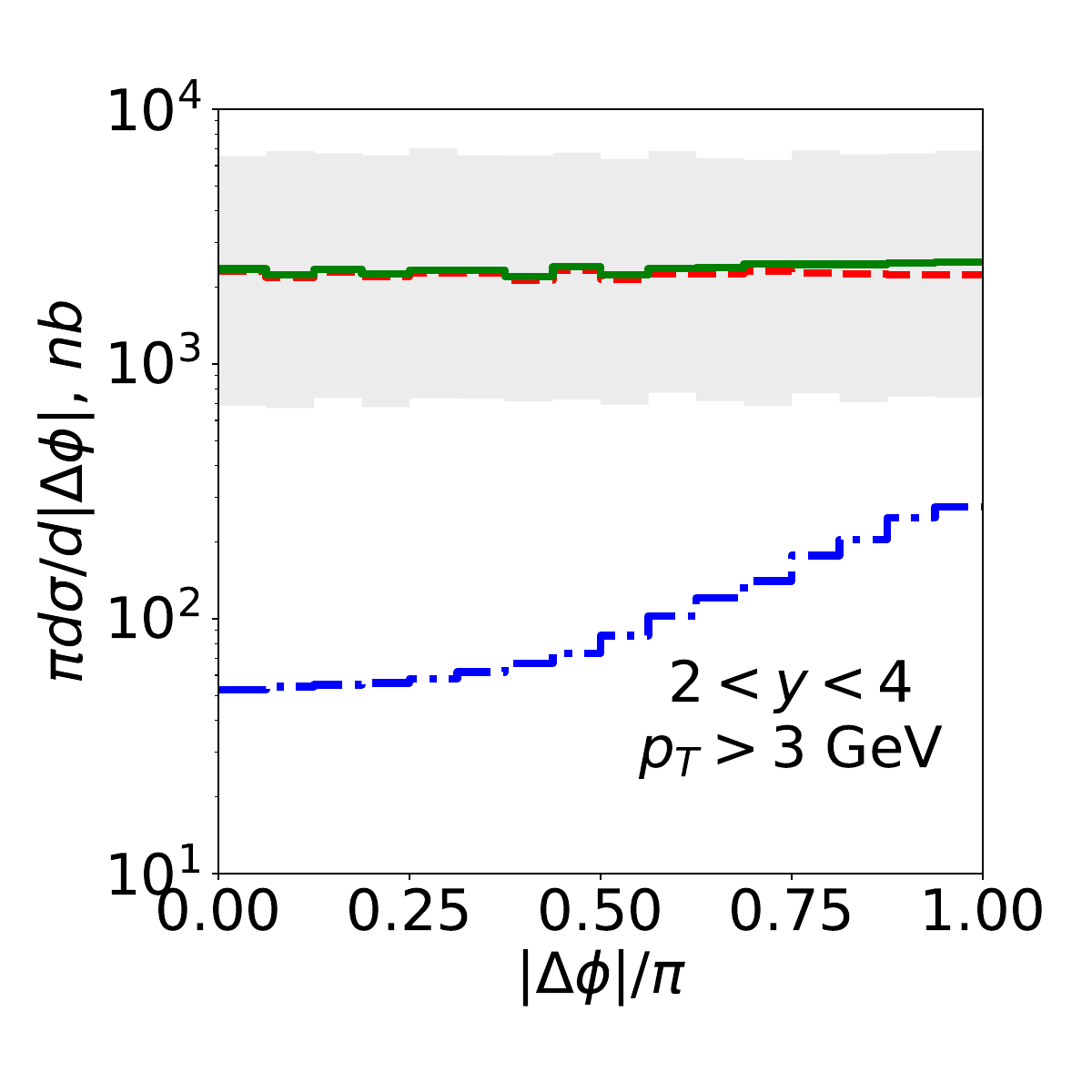}
    \end{minipage}
    \vspace{-5mm}
    \begin{minipage}[b]{0.4\textwidth}
    \includegraphics[width=\textwidth]{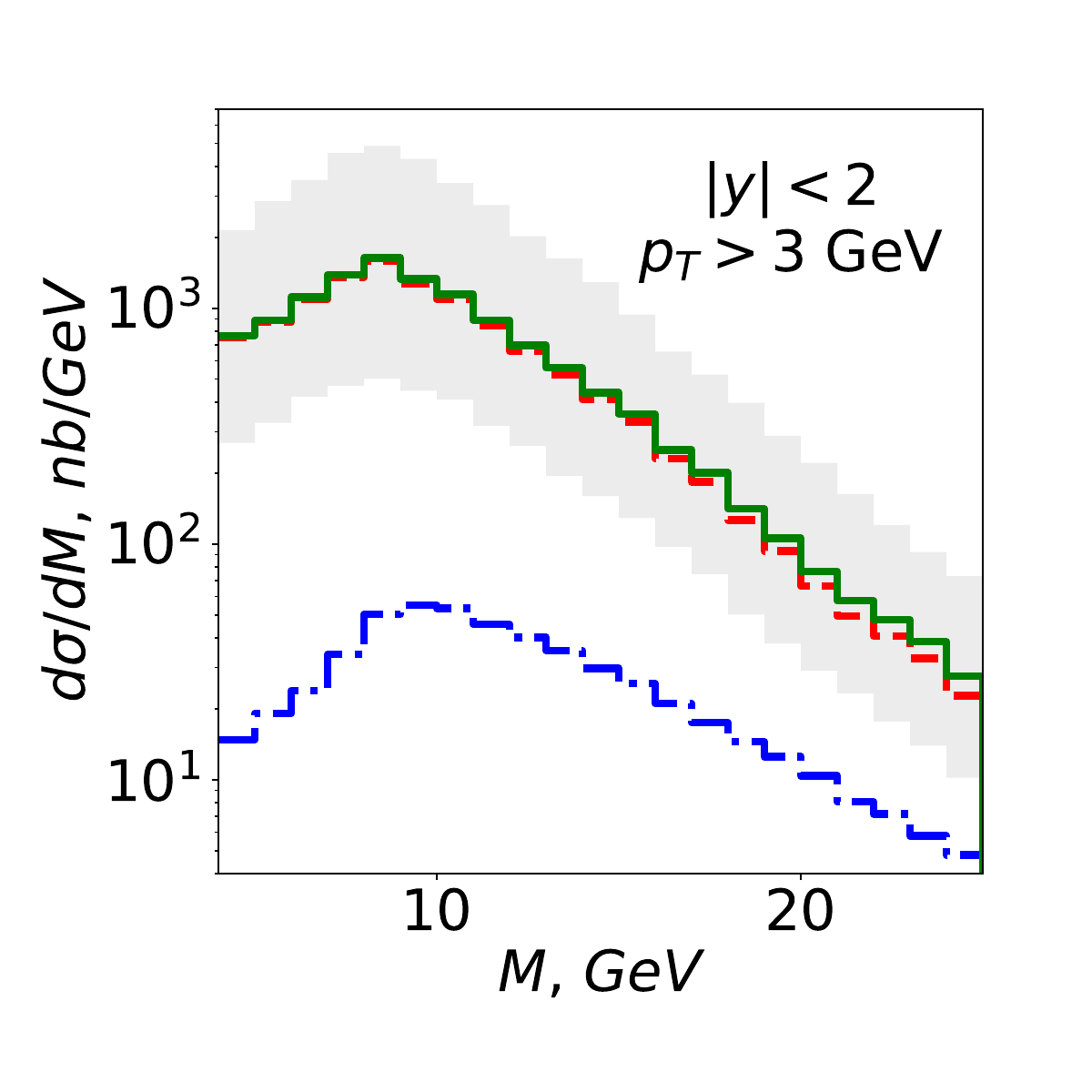}
    \end{minipage}
    \begin{minipage}[b]{0.4\textwidth}
    \includegraphics[width=\textwidth]{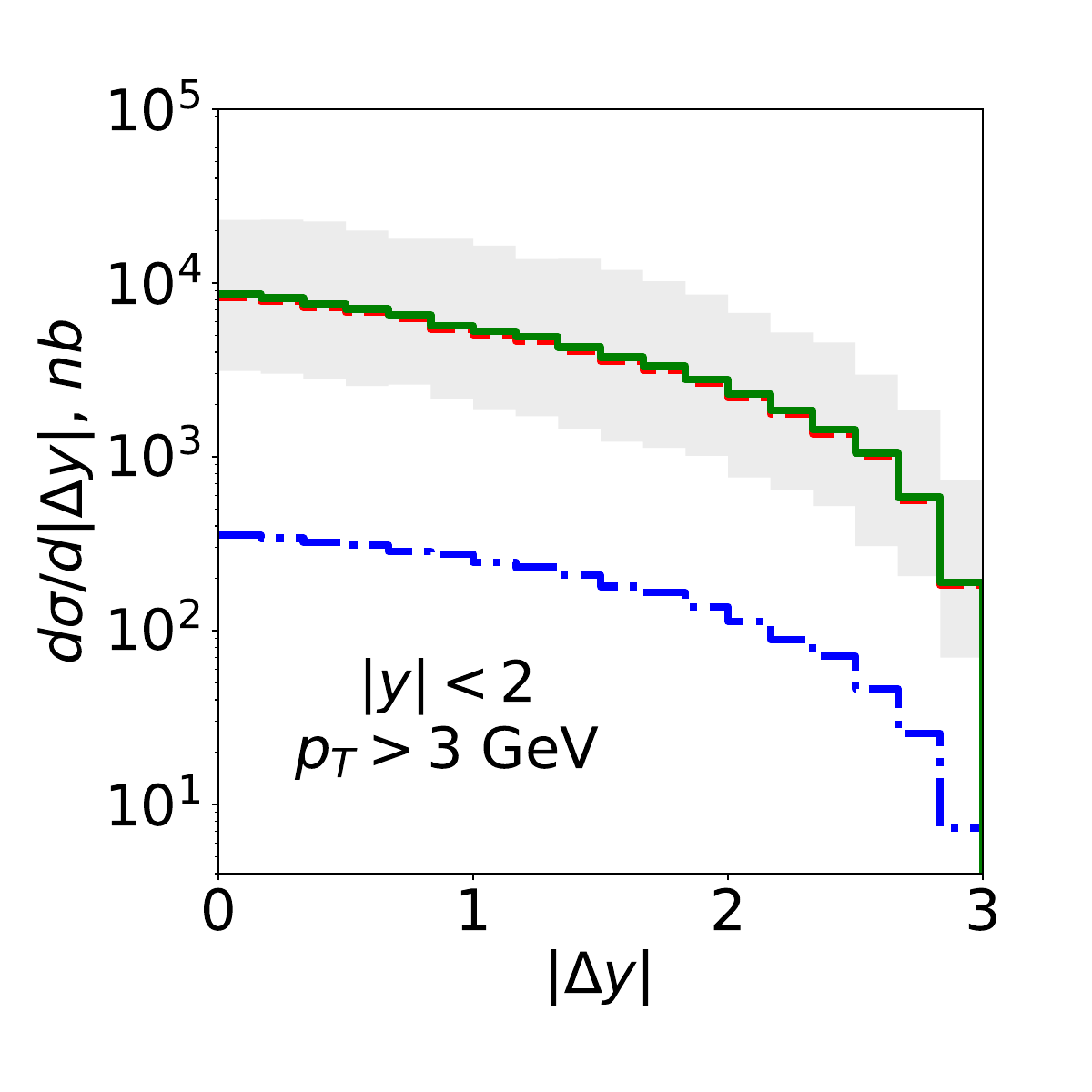}
    \end{minipage}
    \vspace{-5mm}
    \begin{minipage}[b]{0.4\textwidth}
    \includegraphics[width=\textwidth]{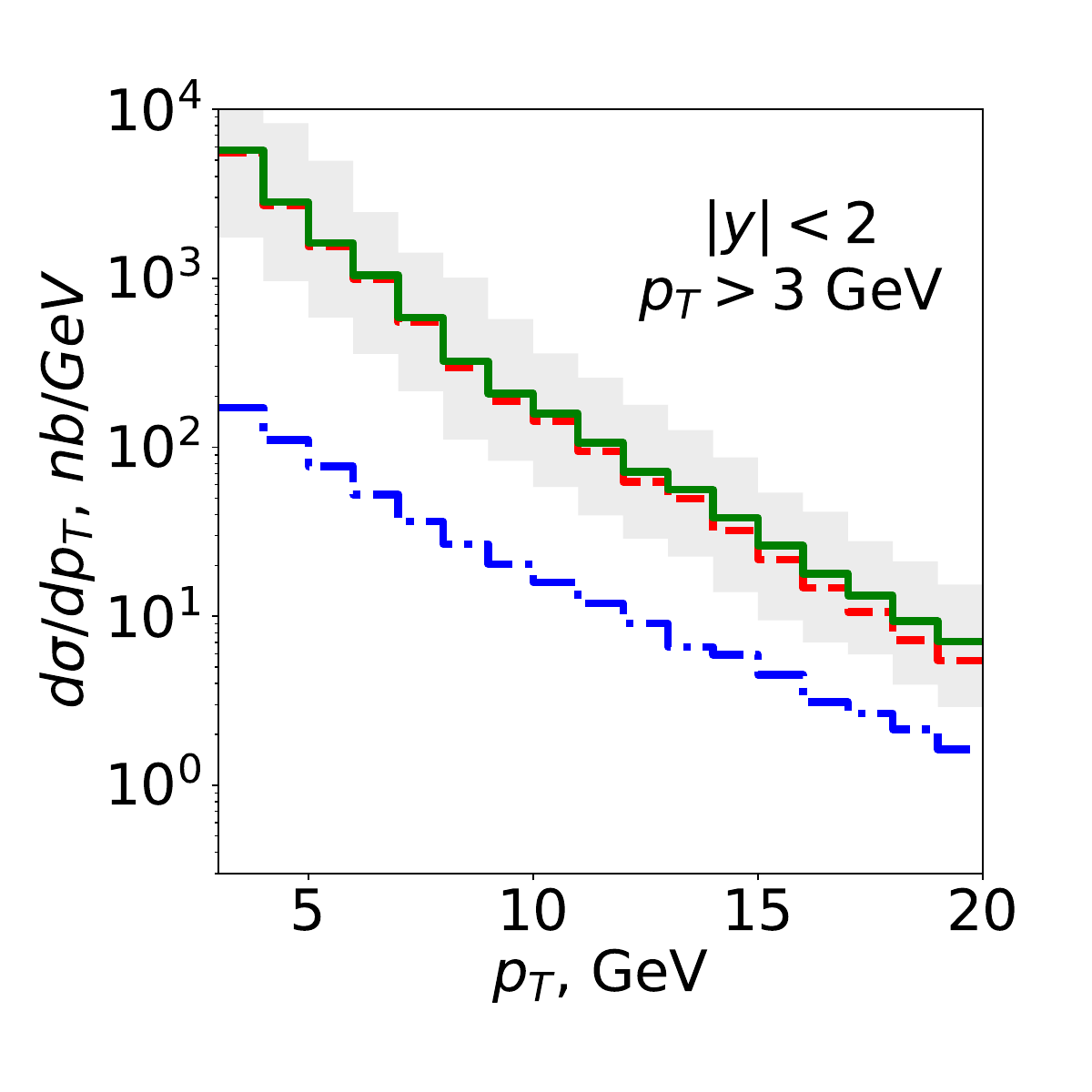}
    \end{minipage}
    \begin{minipage}[b]{0.4\textwidth}
    \includegraphics[width=\textwidth]{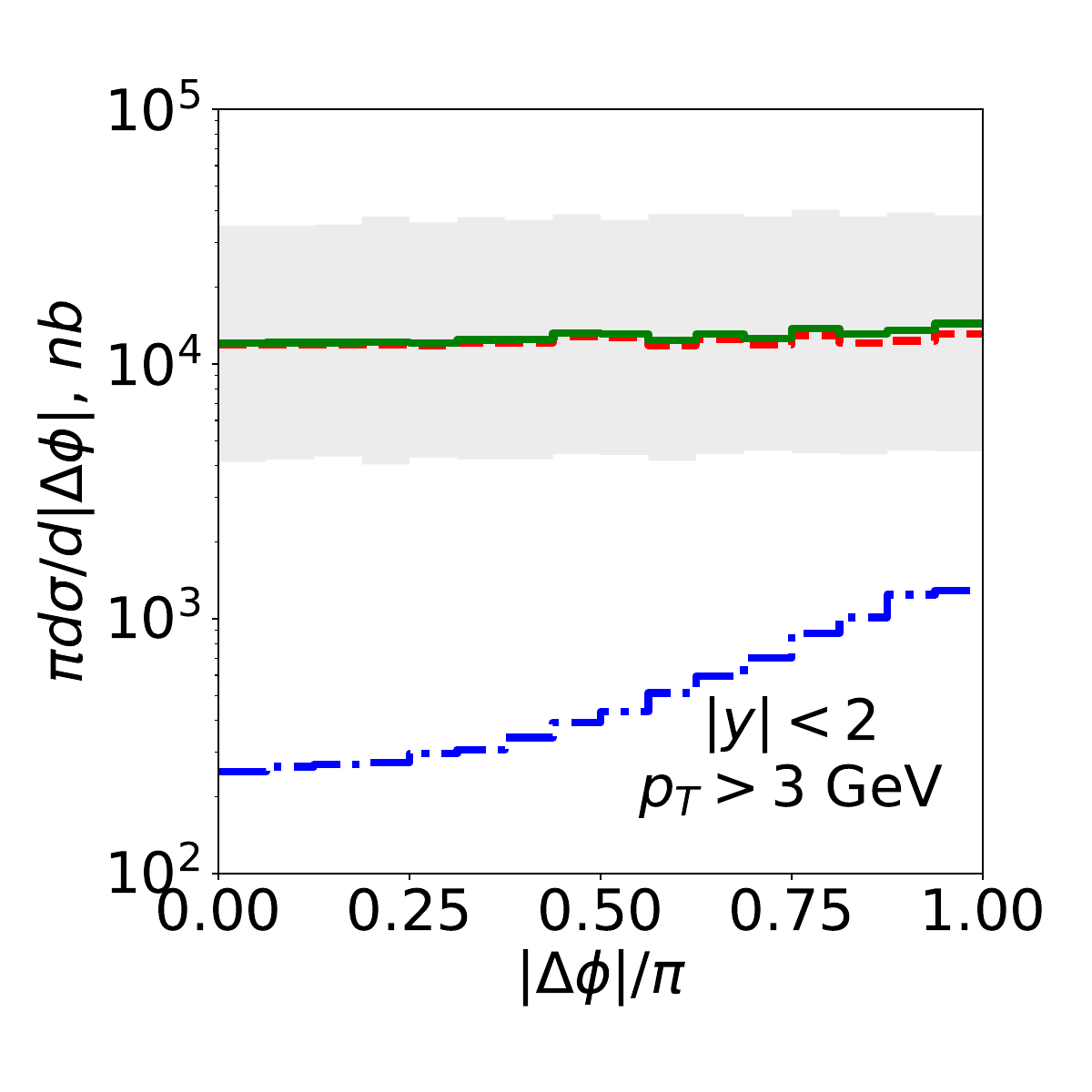}
    \end{minipage}
    \vspace{-1mm}
\label{pic:test_kt}
\caption{Differential cross sections for $D^0D^0$ pair production as functions of the invariant mass $M$, the azimuthal angle difference $|\Delta\phi|/\pi$, the rapidity difference $|\Delta y|$, and the $p_T$ of a single $D$-meson in the forward and central kinematic regions at $\sqrt{s}=13$ TeV. The notation for the curves is the same as in Fig. 1.}
\end{center}
\labelf{fig01}
\vspace{-5mm}
\end{figure}

\bibliographystyle{pepan}
\bibliography{biblio}

\end{document}